\documentclass[prb,aps,twocolumn,, nofootinbib]{revtex4-2} 

\usepackage{amsmath}
\usepackage{amssymb}
\usepackage{mathrsfs}
\usepackage[bbgreekl]{mathbbol}
\usepackage{mathtools}
\usepackage{physics}
\usepackage[american]{babel}

\usepackage{multirow}
\usepackage{hhline}

\usepackage[x11names]{xcolor}
\usepackage[colorlinks=true, citecolor=blue!90!black, linkcolor=blue!90!black, linktocpage=true, urlcolor=red!70!black]{hyperref}

\usepackage{tikz}
\usetikzlibrary{decorations.pathreplacing}

\definecolor{MajBlue}{HTML}{377eb8}
\definecolor{MajRed}{HTML}{e41a1c}

\renewcommand{\eqref}[1]{Eq.~(\ref{#1})}

\makeatletter
\newcommand{\HideAppendixSubsectionsFromToc}{%
  \let\l@subsection\@gobbletwo
  \let\l@subsubsection\@gobbletwo
}
\makeatother



\begin{document}

\title{
Fermionic Villain model with exact lattice chiral symmetries
}

\hfill MIT-CTP/6081

\author{
Zhiyao Lu and
Shu-Heng Shao
}

\affiliation{Center for Theoretical Physics --- a Leinweber Institute, Massachusetts Institute of Technology}

\begin{abstract}

We present a fermionic lattice Hamiltonian that exactly realizes the $\mathrm{U(1)}_\mathrm{L}\times \mathrm{U(1)}_\mathrm{R}$ global symmetry and the associated chiral anomalies of a massless Dirac fermion in 1+1 dimensions. 
The construction couples Villain bosons to a Kitaev chain of Majorana fermions, where the bosonic fields are essential for evading Nielsen–Ninomiya-type no-go theorems.   
For two copies of our model, we realize the anomaly-free $\mathrm{U(1)}_{3450}$ global symmetry and construct symmetric boundary conditions. 
We analytically demonstrate symmetric mass generation by mapping the symmetry-preserving six-fermion interactions to fermion bilinear terms using fermionic T-duality.  
Finally, by gauging general anomaly-free global symmetries, we obtain a broad class of lattice chiral gauge theories. 
As nontrivial applications, we compute the mass spectra of the lattice Schwinger model and the 3450 gauge theory, finding agreement with the corresponding continuum results.
\end{abstract}

\maketitle
\tableofcontents

\section{Introduction}
\label{sec:introduction}

Chiral symmetry plays several central roles in quantum field theory. In quantum chromodynamics, the spontaneous breaking of the chiral global symmetry controls much of low-energy hadron physics: the pions arise as approximate Nambu–Goldstone bosons, and their interactions are organized by the chiral Lagrangian. At much higher energies, chirality is equally fundamental because the Standard Model is a chiral gauge theory, with left- and right-handed fermions transforming differently under the electroweak gauge group.

A microscopic realization of chiral symmetry is both conceptually interesting and practically important. On the lattice, however, such a realization faces a fundamental obstruction. Nielsen and Ninomiya \cite{Nielsen:1980rz,Nielsen:1981xu,Nielsen:1981hk} showed that a locality-preserving, compact, continuous chiral symmetry with an 't Hooft anomaly cannot be realized using non-interacting lattice fermions in the same spacetime dimension. 
See also Refs.~\cite{Friedan:1982nk,Shamir:1993bi}.  
More recently, this no-go theorem was extended to interacting Hamiltonian lattice models in Refs.~\cite{Fidkowski:2023sif,Kapustin:2024rrm,Liu:2026atf}, which showed that a global symmetry with an infinite-order, or equivalently torsion-free, 't Hooft anomaly cannot be realized exactly on a lattice with a finite-dimensional local Hilbert space.

Once continuous bosonic degrees of freedom are introduced, these no-go theorems no longer apply because quantization gives rise to an infinite-dimensional local Hilbert space, much like that of the quantum harmonic oscillator.   
Indeed, continuous Abelian global symmetries with 't Hooft anomalies can be realized exactly in the modified Villain model \cite{Sulejmanpasic:2019ytl,Gorantla:2021svj}, a Euclidean lattice formulation building on the earlier work of Ref.~\cite{Gross:1990ub}.\footnote{See Ref.\ \cite{Baig:2026zgb} for discussions of how related no-go theorems constrain  Euclidean models with continuous bosonic fields.} The modified Villain model was subsequently generalized to the Hamiltonian formalism in both 1+1d \cite{Cheng:2022sgb,Fazza:2022fss} and 3+1d \cite{Lu:2026jnq}. A closely related rotor formalism, based on periodic scalar fields, likewise realizes a variety of global symmetries with perturbative anomalies \cite{Fidkowski:2025rsq,Thorngren:2026ydw}. These developments have, in turn, led to new constructions of lattice chiral gauge theories \cite{Berkowitz:2023pnz,Thorngren:2026ydw,Seifnashri:2026ema,Lu:2026jnq}.

However, these models do not have fermions. 
They realize either the bosonized version of the chiral anomaly in 1+1d \cite{Gorantla:2021svj,Cheng:2022sgb,Fazza:2022fss,Seifnashri:2026ema,Baig:2026zgb}, or the anomaly of a Goldstone boson in a spontaneous-symmetry-breaking phase in 3+1d \cite{Lu:2026jnq}. 
It remains an important open question to construct a \textit{fermionic} lattice model with chiral global symmetries and anomalies.

In this paper, we address this question in 1+1d by presenting a fermionic lattice Hamiltonian, which we call the \textit{fermionic Villain model}, that couples a Kitaev chain of Majorana fermions to the bosonic Villain Hamiltonian. 
This model realizes the $\mathrm{U(1)}_\mathrm{L}\times\mathrm{U(1)}_\mathrm{R}$ chiral global symmetry of a massless Dirac fermion. This is the simplest example of a chiral global symmetry, whose anomaly, first studied by Schwinger \cite{Schwinger:1962tp} (see also Ref.~\cite{Johnson:1963vz}), is characterized by a level $\pm 1$ Chern--Simons term in 2+1d. 
This anomaly is known to be incompatible with a bosonic quantum system, where the levels are constrained to be even integers (see, e.g., Ref.~\cite{Seiberg:2016rsg} for a pedagogical discussion). 
This lattice realization of chiral symmetry allows us to demonstrate anomaly cancellation and to construct and solve chiral gauge theories exactly on the lattice.

Let us compare this new lattice realization of chiral symmetry with those in the staggered fermion model \cite{Banks:1975gq}. In the latter, the $\mathrm{U(1)}_\mathrm{L}\times\mathrm{U(1)}_\mathrm{R}$ symmetry is realized either as the non-Abelian Onsager algebra \cite{Chatterjee:2024gje} or as the noncompact group $\mathrm{U(1)}\times\mathbb{R}$ \cite{Creutz:2001wp,Chatterjee:2024gje}, in accordance with the no-go theorems. By contrast, the lattice chiral symmetry of the fermionic Villain model obeys the ordinary group law of $\mathrm{U(1)}_\mathrm{L}\times\mathrm{U(1)}_\mathrm{R}$ and is preserved even in the presence of four-fermion interactions.

The rest of this paper is organized as follows. Sec.~\ref{sec:fVillain} introduces the fermionic Villain model and presents its exact $\mathrm{U(1)}_\mathrm{L}\times\mathrm{U(1)}_\mathrm{R}$ chiral global symmetry. We demonstrate the associated chiral anomaly by computing the Schwinger term and through spectral flow. We also establish an exact fermionic T-duality on the lattice. 
Sec.~\ref{sec:3450} realizes the exact 3450 global symmetry and constructs a symmetry-preserving boundary condition. We analytically demonstrate symmetric mass generation by mapping six-fermion interactions to bilinear mass terms via fermionic T-duality. Sec.~\ref{sec:gauge} presents a general construction of lattice chiral gauge theories by gauging anomaly-free global symmetries in the fermionic Villain model. As applications, we compute the mass spectra of the Schwinger model and the 3450 gauge theory and compare with the continuum results.

\section{The fermionic Villain model}\label{sec:fVillain}

\subsection{The Hilbert space}

Consider a 1d spatial lattice with $N$ sites. 
We first introduce the fields for the bosonic Villain Hamiltonian model in Refs.  \cite{Cheng:2022sgb,Fazza:2022fss}. 
On every site $j$ there is a real-valued scalar field $\phi_j$ and its conjugate field $p_j$. 
On every link $(j,j+1)$, there is a Villain gauge field $w_{j,j+1}$ and its conjugate field $\tilde\phi_{j,j+1}$. 
They obey the canonical commutation relations:
\begin{align}
[\phi_j , p_{j'} ] =  i\delta_{j,j'},~~~[w_{j,j+1}, \tilde\phi_{j',j'+1}] = -  i \delta_{j,j'}.
\end{align}
In the bosonic Villain model, $w_{j,j+1}$ is integer-valued and satisfies $\exp(2\pi  i w_{j,j+1})=1$. 
It is associated with the gauge transformation $\phi_j\sim \phi_j+2\pi m_j,~w_{j,j+1}\sim w_{j,j+1} -m_{j+1}+m_j$ where $m_j\in \mathbb{Z}$ is the integer gauge parameter. 
This integer gauge transformation, which is implemented by the Gauss law $\exp\left(2\pi i p_j+i\tilde{\phi}_{j-1,j}-i\tilde{\phi}_{j,j+1}\right)=1$, effectively renders the scalar field compact. 
Below we will modify these two local constraints by coupling to fermions. 

We introduce two Majorana fermions $\gamma_{j,j+1}, \gamma'_{j,j+1}$ on every link, obeying the anticommutation relations $\{ \gamma_{j,j+1} , \gamma_{j',j'+1} \} = 2\delta_{j,j'}$,$
\{ \gamma_{j,j+1}' , \gamma_{j',j'+1}' \} = 2\delta_{j,j'}$, and 
$\{\gamma_{j,j+1}, \gamma_{j',j'+1}'\}=0.$ 
While the bosons always satisfy the periodic boundary condition, we allow either periodic (Ramond) or anti-periodic (Neveu-Schwarz) boundary conditions for the fermions, i.e., $\gamma_{N,N+1} = (-1)^\eta \gamma_{0,1},\gamma'_{N,N+1} = (-1)^\eta \gamma'_{0,1}$ with $\eta=0,1$. 

The Hilbert space of our \textit{fermionic Villain model} is subject to the following two Gauss law constraints: 
\begin{subequations}\label{eq:gauss}
\begin{align}
&    \exp\left(i \pi  p_j+\frac{i}{2}\tilde{\phi}_{j-1,j}-\frac{i}{2}\tilde{\phi}_{j,j+1}\right)=i\gamma'_{j-1,j}\gamma_{j,j+1},
\label{eq:gauss1}\\
&  \exp\left(2\pi i w_{j,j+1}\right)
=i\gamma_{j,j+1}\gamma'_{j,j+1}\label{eq:gauss2}.
\end{align}
\end{subequations}
Note that the two fermion bilinears on the right-hand side correspond to the trivial and nontrivial pairings in the Kitaev chain \cite{Kitaev:2000nmw}. 
These Gauss laws were proposed in \cite{Seifnashri:2026ema} in the context of the lattice chiral gauge theory, and can be understood as the gauging of a $\mathbb{Z}_2$ (momentum) symmetry of the bosonic  Villain model using fermions \cite{Pace:2024oys}.
The first Gauss law implements a fermionic Villain gauge transformation:
\begin{equation}
\begin{aligned}
&\phi_j\sim \phi_j + \pi m_j ,~~~
w_{j,j+1}\sim w_{j,j+1} 
 - {m_{j+1}-m_j\over2}
,~~~\\
&\gamma_{j,j+1}\sim (-1)^{m_j}\gamma_{j,j+1},~~~
\gamma'_{j,j+1}\sim (-1)^{m_{j+1}} \gamma'_{j,j+1} ,
\end{aligned}
\end{equation}
with $m_j\in \mathbb{Z}$.  
The second Gauss law states that $w_{j,j+1}$ is an integer (half-integer) if the fermion occupation number $(1-i\gamma_{j,j+1}\gamma'_{j,j+1})/2$ is 0 (1). 
In the continuum, this corresponds to coupling the bosonic theory to a spin TQFT (i.e., the Arf invariant) and then gauging a diagonal $\mathbb{Z}_2$ global symmetry, resulting in a fermionic theory \cite{Alvarez-Gaume:1987wwg}. See Refs.~\cite{Gaiotto:2015zta,Kapustin:2017jrc,Karch:2019lnn,Ji:2019ugf,Lin:2019hks,Hsieh:2020uwb,Kulp:2020iet} for modern discussions on fermionization.

There is a $\mathbb{Z}_2^\text{F}$ fermion parity symmetry  generated by $(-1)^\text{F} = \prod_{j=1}^N i \gamma_{j,j+1} \gamma'_{j,j+1}$. 
Note that $\gamma_{j,j+1}$ and $\gamma_{j,j+1}'$ are not gauge-invariant under the Gauss law constraints in  \eqref{eq:gauss}.
Rather, the simplest fermionic local operators are
\begin{equation}
\begin{aligned}\label{eq:localfermion}
  &  \psi_{\text{L},j}=e^{-i\phi_j-\frac i2\tilde\phi_{j,j+1}} \, \gamma_{j,j+1},\\
  &\psi_{\text{L},j+\frac12}=e^{-i\phi_{j+1}-\frac i2 \tilde\phi_{j,j+1}}\, \gamma'_{j,j+1},
\\
 &  \psi_{\text{R},j}=e^{-i \phi_j+\frac i2\tilde\phi_{j,j+1}}\, \gamma_{j,j+1},\\
  &\psi_{\text{R},j+\frac12}=e^{-i\phi_{j+1}+\frac i2\tilde\phi_{j,j+1}}\, \gamma'_{j,j+1},
\end{aligned}
\end{equation}
and their Hermitian conjugates. 
We will denote the first (last) two operators in \eqref{eq:localfermion} collectively by $\psi_{\text{L},\ell}$  ($\psi_{\text{R},\ell}$) where $\ell$ can be an integer $j$ or a half-integer $j+\frac12$. 
We stress that our fermionic operators are local operators.  
In contrast, the model of  Ref.~\cite{Baig:2026zgb} is bosonic and its fermion operators are non-local operators attached to topological lines.

\subsection{A Hamiltonian for the Dirac fermion}

The simplest Hamiltonian for the fermionic Villain model is
\begin{equation}\label{eq:Hamiltonian}
    H=  
{1\over 2R^2}\sum_{j=1}^N p_j^2
+{R^2\over2}
\sum_{j=1}^N \left( \frac{\phi_{j+1}-\phi_j}{2\pi} +w_{j,j+1}\right)^2 \,.
\end{equation}
Even though this Hamiltonian looks identical to the bosonic model in Refs.~\cite{Cheng:2022sgb,Fazza:2022fss}, it is subject to the fermionic Gauss laws  in \eqref{eq:gauss}. 
The continuum limit of this Hamiltonian is a massless Dirac fermion with a four-fermion Thirring interaction:
\begin{equation}
\begin{aligned}\label{eq:Thirring}
    \mathcal{L}&=i\psi_\text{L}^\dagger (\partial_t+\partial_x)\psi_\text{L}
    +i\psi_\text{R}^\dagger(\partial_t-\partial_x)\psi_\text{R} \\
    &
    -2g(\psi_\text{L}^\dagger \psi_\text{L})(\psi_\text{R}^\dagger \psi_\text{R} )
\end{aligned}
\end{equation}
with $2/R^2 =1+g/\pi$ \cite{Coleman:1974bu}.
We exactly solve the spectrum of the lattice Hamiltonian in App.~\ref{app:spectrum} and show that it matches the continuum spectrum.
In particular, the $R=\sqrt{2}$ model corresponds to a free massless Dirac fermion with $g=0$.

\subsection{U(1)$_\text{L}\times \text{U(1)}_\text{R}$ global  symmetries and chiral anomalies}

The Dirac fermion field theory \eqref{eq:Thirring} has a $[\text{U(1)}_\text{L}\times \text{U(1)}_\text{R}]\rtimes \mathbb{Z}_2^{\cal C}$ global symmetry that acts on the fermion fields as
\begin{subequations}\label{eq:U1LR}
\begin{align}
&\text{U(1)}_\text{L}:~~\psi_\text{L}\to e^{-i \theta} \psi_\text{L} ,~~~&&\psi_\text{R}\to \psi_\text{R},\\
&\text{U(1)}_\text{R}:~~\psi_\text{L}\to  \psi_\text{L} ,~~~&&\psi_\text{R}\to e^{-i \theta}\psi_\text{R},
\end{align}
\end{subequations}
and
\begin{equation}\label{eq:contC}
{\cal C}:~~\psi_\text{L}\leftrightarrow \psi_\text{L}^\dagger,~~~\psi_\text{R}\leftrightarrow \psi_\text{R}^\dagger.
\end{equation}
\eqref{eq:U1LR} is the simplest example of a chiral global symmetry in field theory. The associated 't Hooft anomaly is captured by the 2+1d Chern-Simons terms ${i\over 4\pi }A_\text{L}dA_\text{L} -{i\over 4\pi} A_\text{R}dA_\text{R}$ at level $\pm1$, where $A_\text{L,R}$ are the background gauge fields. 

Our fermionic Villain lattice model realizes an exact, locality-preserving $\text{U(1)}_\text{L}\times \text{U(1)}_\text{R}$ global symmetry. Explicitly, the lattice chiral charges are
\begin{subequations}\label{eq:QLR}
\begin{align}
&Q_\text{L} = \sum_j\left(\frac 12 p_j + w_{j,j+1}\right), \\
&Q_\text{R} = \sum_j\left(\frac 12 p_j -w_{j,j+1}\right).
\end{align}
\end{subequations}
To understand the quantization of the charges, we compute,
\begin{equation}
\begin{aligned}
&e^{2\pi i Q_\text{L} }=e^{2\pi i Q_\text{R} }=e^{ i \pi\sum_j p_j}e^{2\pi i\sum_j w_{j,j+1}}
\\
&= \prod_{j=1}^N i \gamma_{j-1,j}' \gamma_{j,j+1} \prod_{j=1}^N i\gamma_{j,j+1}\gamma_{j,j+1}' = (-1)^{\eta+1},
\end{aligned}
\end{equation}
where we have used \eqref{eq:gauss}, together with the fact that $[p_j ,w_{j',j'+1}]=0$. 
Hence, we learn that $Q_\text{L},Q_\text{R}\in \mathbb{Z}$ for anti-periodic boundary condition ($\eta=1$) and $Q_\text{L},Q_\text{R}\in \mathbb{Z}+\frac12$ for periodic boundary conditions $(\eta=0)$, consistent with the continuum field theory. 
We define the vector and axial charges as
\begin{subequations}
\begin{align}
&Q_\text{V} = Q_\text{L} +Q_\text{R}  = \sum_j p_j ,\\
&Q_\text{A} = Q_\text{L} - Q_\text{R}  =2\sum_j w_{j,j+1}.
\end{align}
\end{subequations}
The model also realizes the $\mathbb{Z}_2^{\cal C}$ charge conjugation symmetry which acts as
\begin{equation}\label{eq:C}
\begin{aligned}
    {\cal C}: ~&\phi_j\mapsto -\phi_j,\ 
    &&p_j\mapsto -p_j,\\
    &\tilde\phi_{j,j+1}\mapsto -\tilde\phi_{j,j+1},\ 
    &&w_{j,j+1}\mapsto -w_{j,j+1},\\
    &\gamma_{j,j+1}\to\gamma_{j,j+1},
    &&\gamma_{j,j+1}'\to\gamma_{j,j+1}'.
    \end{aligned}
\end{equation} 
It acts on the local fermion operators in \eqref{eq:localfermion} as  $\mathcal{C}:~\psi_{\text{L},\ell}\leftrightarrow \psi_{\text{L},\ell}^\dagger,~ \psi_{\text{R},\ell}\leftrightarrow \psi_{\text{R},\ell}^\dagger$, and flips the sign of both $Q_\text{L}$ and $Q_\text{R}$.

The chiral anomaly of the $\mathrm{U(1)}_\mathrm{L}\times \mathrm{U(1)}_\mathrm{R}$ symmetry can be diagnosed from the Schwinger term, i.e., the equal-time commutator of the charge densities \cite{Schwinger:1959xd}. 
The lattice charge densities are
\begin{align}\label{eq:rhoLrhoR}
&\rho_{\text{L},j}=\frac{1}{2}p_j+\frac{\phi_{j+1}-\phi_j}{2\pi}+w_{j,j+1},\\ &\rho_{\text{R},j}=
\frac{1}{2}p_{j+1}-\frac{\phi_{j+1}-\phi_{j}}{2\pi}-w_{j,j+1}, 
\end{align}
which satisfy $Q_\text{L}= \sum_j \rho_{\text{L},j}$ and $Q_\text{R}= \sum_j \rho_{\text{R},j}$. 
The improvement terms $\frac{\phi_{j+1}-\phi_{j}}{2\pi}$ do not affect the total charges, but are essential for the gauge invariance of the charge densities  under the Gauss laws in \eqref{eq:gauss}. 
Their equal-time commutators are
\begin{equation}
\begin{aligned}\label{eq:LRSchwinger}
    &[\rho_{\text{L},j},\rho_{\text{L},j'}]=\frac{i}{4\pi}(\delta_{j+1,j'}-\delta_{j-1,j'}),\\
    &[\rho_{\text{R},j},\rho_{\text{R},j'}]=-\frac{i}{4\pi}(\delta_{j+1,j'}-\delta_{j-1,j'}),\\&[\rho_{\text{L},j},\rho_{\text{R},j'}]=0,
\end{aligned}
\end{equation}
which matches the Schwinger term in the continuum (see App. \ref{app:Schwinger}). 
The charge densities $\rho_{\text{L},j}$ do not commute with each other, signaling an obstruction to imposing mutually commuting Gauss laws. 
This obstruction reflects the 't Hooft anomaly of $\mathrm{U(1)}_\mathrm{L}$, and the same reasoning applies to $\mathrm{U(1)}_\mathrm{R}$.

The local lattice fermionic operators $\psi_{\text{L},j}$ and $\psi_{\text{L},j+\frac12}$ have charges $Q_\text{L}=  -1,Q_\text{R}=0$, and $\psi_{\text{R},j}$ and $\psi_{\text{R},j+\frac12}$ have charges $Q_\text{L}=  0,Q_\text{R}=-1$:
\begin{subequations}
\begin{align}
&[Q_\text{L}, \psi_{\text{L},\ell}] = - \psi_{\text{L},\ell},&&[Q_\text{R}, \psi_{\text{L},\ell}] = 0,\\
&[Q_\text{L}, \psi_{\text{R},\ell}] = 0,&&[Q_\text{R}, \psi_{\text{R},\ell}] = - \psi_{\text{R},\ell}.
\end{align}
\end{subequations}
The lattice operators in \eqref{eq:localfermion} are therefore identified with the continuum fermion fields $\psi_\text{L}(t,x)$ and $\psi_\text{R}(t,x)$. 
The complex Dirac mass term is 
\begin{equation}\label{eq:mass}
m \psi_{\text{L},j}^\dagger \psi_{\text{R},j} +h.c. 
= m  \exp(i  \tilde\phi_{j,j+1})+h.c.,
\end{equation}
which has $Q_\text{V}=0$ and $Q_\text{A}=2$. 
Adding this mass term to the Hamiltonian in \eqref{eq:Hamiltonian}  trivially gaps the system.

The anomaly can also be diagnosed from spectral flow \cite{Schwimmer:1986mf}; see Refs.~\cite{Lin:2019kpn,Benjamin:2020swg,Cheng:2022sgb,Pace:2024oys,Thorngren:2026ydw,Lew-Smith:2026jqb} for recent discussions. 
Consider a $2\pi$ U(1)$_\text{L}$ rotation in a segment on the lattice from site $j=j_1$ to site $j=j_2$:
\begin{equation}
\exp(2\pi i \sum_{j=j_1}^{j_2} \rho_{\text{L},j} )
= i (-1)^{j_2-j_1+1} \psi_{\text{L},j_1-\frac12}
\psi^\dagger_{\text{L},j_2+\frac12},
\end{equation}
where we have used \eqref{eq:gauss}. 
Hence, a $2\pi$ flux of U(1)$_\text{L}$ ``pumps" a local fermion with unit  charge, matching the continuum spectral-flow result reviewed in App.~\ref{app:spectralflow}.

\subsection{Fermionic T-duality and Majorana translations}\label{sec:Tduality}

The continuum Dirac fermion field theory in \eqref{eq:Thirring} enjoys a fermionic T-duality \cite{Karch:2019lnn,Ji:2019ugf}, which differs from the usual bosonic T-duality \cite{Giveon:1994fu} by a factor of 2. 
The bosonic T-duality, which maps $R\to 1/R$, is realized exactly on the lattice in the modified Villain lattice model \cite{Gorantla:2021svj,Cheng:2022sgb,Fazza:2022fss}.\footnote{In our convention for the bosonic T-duality, the self-dual radius is $R=1$, which has an enhanced $\mathfrak{su}(2)_1$ current algebra in the continuum CFT. In contrast, the self-dual radius $R=\sqrt{2}$ for the fermionic T-duality is the free Dirac fermion point with a $\mathfrak{u}(1)_4$ current algebra.} 
Our fermionic Villain model similarly has a fermionic T-duality transformation:
\begin{equation}\label{eq:fermionTduality}
\begin{aligned}
    {\cal T}:~&\phi_j\mapsto \frac{1}{2}\tilde\phi_{j,j+1},
    ~~~~~&&\tilde\phi_{j,j+1}\mapsto 2\phi_{j+1},\\
    &q_j \mapsto 2w_{j,j+1},~~~~~&&w_{j,j+1}\mapsto \frac{1}{2}q_{j+1},\\
    &\gamma_{j,j+1}\mapsto \gamma'_{j,j+1},
    ~~~~~&&\gamma'_{j,j+1}\mapsto \gamma_{j+1,j+2},
\end{aligned}
\end{equation}
where we have defined 
\begin{equation}
q_j=p_j + {\tilde \phi_{j-1,j}-\tilde\phi_{j,j+1}\over2\pi}.
\end{equation}
This transformation maps the Hamiltonian in \eqref{eq:Hamiltonian} with parameter $R$ to that at $2/R$ and preserves the Gauss laws in \eqref{eq:gauss}. 
It acts on the charges as $\mathcal{T}:~Q_\text{L}\to Q_\text{L} ,~Q_\text{R}\to -Q_\text{R}$, and on the local fermions as 
\begin{equation}\label{eq:TonPsi}
\begin{split}
    {\cal T}:~ 
    &\psi_{\text{L},\ell}\mapsto \psi_{\text{L},\ell+\frac12},\ \psi_{\text{L},\ell}^\dagger\mapsto \psi_{\text{L},\ell+\frac12}^\dagger ,\\
    &\psi_{\text{R},\ell}\mapsto \psi_{\text{R},\ell+\frac12}^\dagger,\ \psi_{\text{R},\ell}^\dagger\mapsto \psi_{\text{R},\ell+\frac12}.
\end{split}
\end{equation}
Importantly, the T-duality transformation involves a translation of the Majorana fermions, and $\mathcal{T}^2$ implements a lattice translation by one site.

The self-dual point $R=\sqrt{2}$ corresponds to the free Dirac fermion theory with vanishing Thirring coupling $g=0$. 
The continuum field theory at that point has an enhanced discrete chiral symmetry $\mathbb{Z}_2^{{\cal C}_\text{R}}$ generated by 
\begin{equation}
\mathcal{C}_\text{R} :~~
 \psi_\text{L}\to \psi_\text{L},~~~\psi_\text{L}^\dagger\to \psi_\text{L}^\dagger,~~~\psi_\text{R}\leftrightarrow \psi_\text{R}^\dagger.
\end{equation}
This symmetry flips the sign of a single Majorana-Weyl fermion, and has a mod 8 anomaly \cite{2013NJPh...15f5002Q,2012PhRvB..85x5132R,Gu:2013azn,Kapustin:2014dxa,Tong:2019bbk,Seiberg:2023cdc,BoyleSmith:2024qgx}.\footnote{In contrast, the $\mathbb{Z}_2^\text{L}$ transformation $e^{i \pi Q_\text{L}}$, which acts as an order-two transformation on local fields, has a mod 4 anomaly of the type in Ref.~\cite{Gu:2012ib}.}
The total internal symmetry is enhanced to $\text{O(2)}_\text{L}\times \text{O(2)}_\text{R}$. 
From \eqref{eq:TonPsi}, it is clear that the discrete chiral symmetry $\mathcal{C}_\text{R}$ in the continuum  emanates \cite{Cheng:2022sgb} from the unitary symmetry $\cal T$, which involves a Majorana translation. 
However, they obey different algebras, i.e., $\mathcal{C}_\text{R}^2=1$ while $\mathcal{T}^{2N}=1$. 
This is similar to the emanant chiral symmetry from the Kitaev chain \cite{Seiberg:2023cdc}. 
There is a similar discrete chiral symmetry $\mathcal{C}_\text{L} = \mathcal{C}\mathcal{C}_\text{R}$ acting on the left movers, which emanates from $\mathcal{CT}$ on the lattice.

\section{The 3450 chiral global symmetry}\label{sec:3450}

\subsection{Anomaly cancellation}

We have seen that in the fermionic Villain model, U$(1)_\text{L}$ and U$(1)_\text{R}$ are individually anomalous. 
By taking multiple copies of the model, there are various anomaly-free subgroups. 
Consider the U(1) global symmetry whose charge  is:
\begin{equation}\label{eq:Nfrho}
\begin{aligned}
    &Q=\sum_j\rho_j,~~~\rho_j=\sum_{I=1}^{N_f}\left(n_\text{L}^{(I)}\rho^{(I)}_{\text{L},j}+n_\text{R}^{(I)}\rho^{(I)}_{\text{R},j}\right).
    \end{aligned}
\end{equation}
Here $n^{(I)}_{\text{L,R}}$ are integers, and $N_f$ is the number of flavors. 
This symmetry  is anomaly-free and can be gauged if the lattice Schwinger terms vanish, i.e.,
\begin{equation}\label{eq:anomalycancel}
\begin{aligned}
[\rho_j, \rho_{j'}]&
= {i\over4\pi}
\left(\sum_{I=1}^{N_f}\left(n_\text{L}^{(I)}\right)^2-\left(n_\text{R}^{(I)}\right)^2\right)
(\delta_{j+1,j'} -\delta_{j-1,j'})
\\
&=0,
\end{aligned}
\end{equation} 
which matches the anomaly cancellation condition in the continuum \cite{Halliday:1985tg}.

A well-known example of an anomaly-free global symmetry is the so-called $\mathrm{U(1)}_{3450}$ symmetry introduced in Refs.~\cite{Wen:2013ppa,Wang:2013yta,Wang:2018ugf} (see also Refs.~\cite{Eichten:1985ft,Chen:2012di}). It corresponds to $N_f=2,
n_\text{L}^{(1)}=3,
n_\text{L}^{(2)}=4,
n_\text{R}^{(1)}=5,
n_\text{R}^{(2)}=0,$ 
namely,
\begin{equation}
    Q_{3450}
    =3Q_\text{L}^{(1)}
    +4Q_\text{L}^{(2)}
    +5Q_\text{R}^{(1)},
\end{equation}
where $Q_\text{L,R}$ are defined in \eqref{eq:QLR}. The lattice charge operator $Q_{3450}$ is integer-valued, i.e., $Q_{3450}\in\mathbb{Z}$, and therefore generates a compact $\mathrm{U(1)}_{3450}$ lattice global symmetry. This should be contrasted with the staggered fermion model, where the 3450 symmetry is realized instead as the noncompact group $\mathbb{R}$  \cite{Xu:2025hfs}. 
Below we address two natural questions associated with this anomaly-free global symmetry in the fermionic Villain setting: symmetric mass generation and symmetric boundary conditions.

\subsection{Symmetric mass generation}\label{sec:SMG}

The presence of a 't Hooft anomaly forbids a trivially gapped phase. The inverse question is more nontrivial: given an anomaly-free global symmetry, is there a symmetry-preserving deformation that drives the system to a trivially gapped phase? 
This question is particularly interesting when there is no symmetric, bilinear mass term. Symmetric mass generation \cite{Wen:2013ppa} then refers to the phenomenon in which a gapless theory acquires a mass gap through interactions while preserving an anomaly-free global symmetry. 
See Ref.~\cite{Wang:2022ucy} for a review.

In the continuum, the lightest $\mathrm{U(1)}_{3450}$-symmetric deformations of two massless Dirac fermions are two irrelevant six-fermion interactions \cite{Wang:2013yta},\footnote{See \cite{Tong:2021phe} for an alternative proposal for symmetric mass generation in the 3450 model using gauge interactions.}  which have been numerically studied in Ref.\ \cite{Zeng:2022grc}.  
Below, we analytically demonstrate that the same six-fermion interactions, formulated in the fermionic Villain model, generate a symmetric mass gap. The key step is to map these six-fermion interactions to a fermion bilinear mass term via the fermionic T-duality developed in Sec.~\ref{sec:Tduality}.

In our model, these six-fermion terms correspond to
\begin{subequations}\label{eq:sixfermion}
\begin{align}
    &\left(\psi_{\text{L},j}^{(1)}\right)^\dagger
    \left(\psi_{\text{R},j}^{(1)}\right)^\dagger
    \psi_{\text{L},j}^{(2)}
    \psi_{\text{L},j+\frac12}^{(2)}
    \left(\psi_{\text{R},j}^{(2)}\right)^\dagger
    \left(\psi_{\text{R},j+\frac12}^{(2)}\right)^\dagger\nonumber
    \\&=-e^{-2i\tilde\phi_{j,j+1}^{(2)}+2i\phi_{j}^{(1)}},\\
    &\psi_{\text{L},j-\frac12}^{(1)}
    \psi_{\text{L},j-1}^{(1)}
    \psi_{\text{L},j}^{(2)}
    \left(\psi_{\text{R},j-\frac12}^{(1)}\right)^\dagger
    \left(\psi_{\text{R},j-1}^{(1)}\right)^\dagger
    \psi_{\text{R},j}^{(2)}\nonumber
    \\&=-e^{-2i\tilde\phi_{j-1,j}^{(1)}-2i\phi_{j}^{(2)}},
\end{align}
\end{subequations}
and their Hermitian conjugates. 
These two terms preserve not only U(1)$_\text{3450}$, but  also the larger anomaly-free symmetry U(1)$_\text{3450}\times \text{U(1)}_\text{0543}$, where U$(1)_\text{0543}$ is generated by the following charge
\begin{align}
    Q_{0543}=5Q_\text{L}^{(2)}+4Q_\text{R}^{(1)}+3Q_\text{R}^{(2)}.
\end{align}

Adding the Hermitian conjugates of \eqref{eq:sixfermion}, the six-fermion terms become the following deformations in the Hamiltonian 
\begin{align}
\Delta H =
&-\lambda\sum_j \cos(2\phi_j^{(1)} - 
2\tilde\phi^{(2)}_{j,j+1})\\
&
-\lambda\sum_j
\cos(2\phi_j^{(2)} +2\tilde\phi^{(1)}_{j-1,j}).\nonumber
\end{align}
To see that these two terms drive the system to a gapped phase, we perform the following 
fermionic T-duality transformation:
\begin{equation}
\begin{aligned}\label{eq:SMGTduality}
    &\Phi^{(I)}_j=\frac{1}{2}\tilde\phi_{j-1,j}^{(I)},&&W_{j,j+1}^{(I)}=\frac{1}{2}q_j^{(I)}\\
    &\tilde\Phi^{(1)}_{j,j+1}=2\phi_j^{(1)}-2\tilde\phi_{j,j+1}^{(2)},&&\tilde\Phi^{(2)}_{j,j+1}=2\phi_j^{(2)}+2\tilde\phi_{j-1,j}^{(1)},\\
    &Q_j^{(1)}=2w_{j-1,j}^{(1)}-2q_j^{(2)},&&Q_j^{(2)}=2w_{j-1,j}^{(2)}+2q_{j-1}^{(1)},\\
    &\Gamma^{(I)}_{j,j+1}=(\gamma')^{(I)}_{j-1,j},&&(\Gamma')^{(I)}_{j,j+1}=\gamma^{(I)}_{j,j+1} .
\end{aligned}
\end{equation}
This transformation preserves the Gauss laws in \eqref{eq:gauss} but simplifies the six-fermion deformations to
\begin{equation}
\begin{aligned}
\Delta H=&-\lambda\sum_j \cos(\tilde\Phi_{j,j+1}^{(1)})
-\lambda\sum_j
\cos(\tilde\Phi_{j,j+1}^{(2)} )\\
=&- {\lambda\over2} \sum_j\left[(\Psi_{\text{L},j}^{(1)})^\dagger\Psi_{\text{R},j}^{(1)}
+(\Psi_{\text{L},j}^{(2)})^\dagger 
\Psi_{\text{R},j}^{(2)}\right]+h.c.,
\end{aligned}
\end{equation}
where the $\Psi$'s are lattice fermionic local operators defined as in \eqref{eq:localfermion} for the new fields. 
(For example, $\Psi_{\text{L},j}=e^{-i\Phi_j -\frac i2 \tilde\Phi_{j,j+1}}\Gamma_{j,j+1}$.)
The six-fermion terms in the original duality frame become a literal mass term in \eqref{eq:mass} in the new T-dual frame,  consistent with the continuum analysis in Refs.~\cite{Wang:2013yta,Wang:2018ugf}. 
This makes it clear that, as $\lambda\to \infty$, the system is driven to a trivially gapped phase. 
In App.\ \ref{app:SMG}, we give a more detailed explanation.

\subsection{Symmetric boundary conditions}

If a theory has a global symmetry $G$ with an 't Hooft anomaly, then it does not admit a $G$-symmetric, simple boundary condition (see, e.g., Refs.~\cite{Wang:2013yta,Cho:2016xjw,Han:2017hdv,Jensen:2017eof,Numasawa:2017crf,Smith:2019jnh,Smith:2020rru,Smith:2020nuf,Thorngren:2020yht,Choi:2023xjw,Wei:2025zyd}).\footnote{A boundary condition is called simple if the ground state is nondegenerate when it is imposed at both ends of an interval \cite{Choi:2023xjw}. A simple Cardy boundary state cannot be written as the sum of other Cardy boundary states.} Indeed, we will construct a two-parameter family of lattice boundary conditions preserving  the anomaly-free 3450 global symmetry in the fermionic Villain formalism. Such symmetric boundary conditions in the continuum have been discussed in Refs.\ \cite{Smith:2019jnh,vanBeest:2023dbu,Arias-Tamargo:2026urw,Antinucci:2026uuh,Wei:2026fsn}, with applications to the $s$-wave sector of the monopole-electron scattering process in 3+1d \cite{Affleck:1993np,Maldacena:1995pq}. 
In fact, these boundary conditions are invariant under the larger U(1)$_\text{3450}\times \text{U(1)}_\text{0543}$ symmetry. 

Consider a spatial lattice that starts from link $(0,1)$ and extends to the right. We place two copies of the fermionic Villain model on this lattice, but we omit the fields on link $(0,1)$ in the second copy. We impose the following boundary conditions,
\begin{equation}\label{eq:bc}
    \phi_1^{(1)}-\tilde\phi_{12}^{(2)}=\varphi_1,\ \phi_{1}^{(2)}+\tilde\phi_{01}^{(1)}=\varphi_2,
\end{equation}
where $\varphi_1,\varphi_2$ are real numbers labeling the boundary conditions. 
For the first copy, the boundary includes link 01 and site 1, while for the second copy, the boundary includes site 1 and link $(1,2)$. 

Away from the boundary, we impose the original Gauss laws,
\begin{equation}
\begin{aligned}
    &i(\gamma')^{(I)}_{j-1,j}\gamma^{(I)}_{j,j+1} \exp\left(i\pi p^{(I)}_{j}+\frac{i}{2}\tilde{\phi}^{(I)}_{j-1,j}-\frac{i}{2}\tilde{\phi}^{(I)}_{j,j+1}\right)=1,\\
    &i\gamma^{(I)}_{j,j+1}(\gamma')^{(I)}_{j,j+1}e^{2\pi i w^{(I)}_{j,j+1}}=1,\\
    &i\gamma^{(1)}_{12}(\gamma')^{(1)}_{12}e^{2\pi i w^{(1)}_{12}}=1,
\end{aligned}
\end{equation}
where $j\ge 2$. On the boundary, we impose the following Gauss laws
\begin{equation}
    \begin{split}
        &\exp\left(2\pi i w_{12}^{(2)}+2\pi i p_1^{(1)}-i\tilde\phi_{12}^{(1)}+i\tilde\phi_{01}^{(1)}\right)i\gamma_{12}^{(2)}(\gamma')_{12}^{(2)}=1,\\
        &\exp\left(2\pi i w_{01}^{(1)}-2\pi i p_1^{(2)}+i\tilde\phi_{12}^{(2)}\right)i\gamma_{01}^{(1)}(\gamma')_{01}^{(1)}=1.
    \end{split}
\end{equation}
These Gauss laws preserve the boundary conditions \eqref{eq:bc}. 

By taking products of the Gauss laws, we find the following quantized and gauge-invariant charges
\begin{equation}
    \begin{split}
        Q_{3450}=\sum_{j\ge 1}&\left(4p^{(1)}_j-2w^{(1)}_{j,j+1}+2p^{(2)}_j+4w^{(2)}_{j,j+1}\right)\\
        &-2w_{01}^{(1)}+\frac{2}{\pi}\tilde\phi_{01}^{(1)},
    \end{split}
\end{equation}
\begin{equation}
    \begin{split}
        Q_{0543}=\sum_{j\ge 1}&\left(2p^{(1)}_j-4w^{(1)}_{j,j+1}+4p^{(2)}_j+2w^{(2)}_{j,j+1}\right)\\
        &-4w_{01}^{(1)}+\frac{1}{\pi}\tilde\phi_{01}^{(1)}.
    \end{split}
\end{equation}
The existence of these charges proves that our boundary conditions are invariant under U(1)$_\text{3450}\times \text{U(1)}_\text{0543}$.

Furthermore, we can find a Hamiltonian that commutes with the boundary conditions, Gauss laws and the symmetry charges. This Hamiltonian should contain bulk terms in \eqref{eq:Hamiltonian} starting from site 2. The boundary terms can be chosen to be:
\begin{equation}
\begin{split}
    H_{\text{boundary}}=&\lambda_1\left(w^{(1)}_{01} - p^{(2)}_{1}+\frac{\phi_1^{(1)}}{2\pi}\right)^2\\
    +&\lambda_2\left(w^{(2)}_{12} + p^{(1)}_{1}+\frac{\phi_2^{(2)}-\phi_1^{(2)}}{2\pi}\right)^2.
\end{split}
\end{equation}
Using the fermionic T-duality transformation \eqref{eq:SMGTduality}, one can explicitly verify that these two boundary terms remove the infinite degeneracy of the edge modes, leading to a well-defined spectrum.

\section{Lattice chiral gauge theories}\label{sec:gauge}

\subsection{Gauging anomaly-free chiral symmetries}\label{sec:chiralgauge}

We start with $N_f$ copies of the fermionic Villain Hamiltonian \eqref{eq:Hamiltonian} and consider the anomaly-free U(1) global symmetry generated by $Q$ in \eqref{eq:Nfrho}. 
To gauge $Q$, we need to express it in terms of the charge density. 
There are two different charge densities: the gauge-invariant charge densities in \eqref{eq:rhoLrhoR}, and the  charge densities that commute with the Hamiltonian. Specifically, the latter are
\begin{subequations}
\begin{align}\label{eq:rhoLrhoRhat}
&\hat\rho_{\text{L},j}=\frac{1}{2}p_j+\frac{\tilde\phi_{j-1,j}-\tilde\phi_{j,j+1}}{4\pi}+w_{j,j+1},\\ &\hat\rho_{\text{R},j}=
\frac{1}{2}p_{j+1}+\frac{\tilde\phi_{j,j+1}-\tilde\phi_{j+1,j+2}}{4\pi}-w_{j,j+1}.
\end{align}
\end{subequations}
These charge densities give the same total charges as \eqref{eq:rhoLrhoR}, but the $\hat\rho$'s do not commute with the Gauss laws in \eqref{eq:gauss}.
For later convenience, we use the charge densities that commutes with the Hamiltonian and write 
$Q=\sum_j\hat\rho_j$ and $\hat\rho_j=\sum_I\left(n_{\text{L}}^{(I)}\hat\rho^{(I)}_{\text{L},j}+n_{\text{R}}^{(I)}\hat\rho^{(I)}_{\text{R},j}\right)$.

An anomaly-free global symmetry can be gauged on the lattice by following the three-step procedure of Refs.~\cite{Lu:2026jnq,Seifnashri:2026ema}, which builds on the framework of Ref. \cite{Seifnashri:2023dpa}. First, the theory is coupled to background gauge fields. Second, we impose Gauss law constraints. Third, we add the kinetic terms for the gauge fields. Importantly, the symmetry need not be disentangled into an onsite form before it can be gauged in our formalism.

To be more specific, we conjugate the system by $U=\exp\left(-i\sum_j \alpha_j \hat\rho_j\right)$, where $A_{j,j+1}=\alpha_{j+1}-\alpha_j$ is the background gauge field. 
Here we have chosen to work with $\hat{\rho}_j$ which commutes with the Hamiltonian but not with the Gauss laws. As a result, the background gauge field $A_{j,j+1}$ appears only in the Gauss laws and not in the Hamiltonian. Equivalently, we can work with the gauge-invariant charge density, and the gauge field would appear in the Hamiltonian instead of the Gauss law. The two presentations are related by a unitary transformation.

After we couple the theory to background gauge fields, the local constraints in \eqref{eq:gauss} become
\begin{equation}\label{eq:gaugeguass1}
    \begin{aligned}
    i(\gamma')^{(I)}_{j-1,j}&\gamma^{(I)}_{j,j+1} \exp\left(\pi i p^{(I)}_{j}+\frac{i}{2}\tilde{\phi}^{(I)}_{j-1,j}-\frac{i}{2}\tilde{\phi}^{(I)}_{j,j+1}\right)\\
    &=\exp\left(i\frac{n_\text{R}^{(I)}-n_\text{L}^{(I)}}{2}A_{j-1,j}\right)\,,
\end{aligned}
\end{equation}
\begin{equation}\label{eq:gaugeguass2}
\begin{aligned}
    i\gamma^{(I)}_{j,j+1}&(\gamma')^{(I)}_{j,j+1}\exp\left(2\pi i w^{(I)}_{j,j+1}\right)=\\
    &\exp\left(-i\frac{n_\text{L}^{(I)}}{2}A_{j,j+1}-i\frac{n_\text{R}^{(I)}}{2}A_{j-1,j}\right)\,.
\end{aligned}
\end{equation}

We then promote $A_{j,j+1}$ to an operator and introduce its conjugate momentum $E_{j,j+1}$, satisfying the following canonical commutation relation $[A_{j,j+1},E_{j',j'+1}]=i\delta_{j,j'}$. We impose the Gauss law associated with the U(1) gauge group:
\begin{equation}\label{eq:gaugeguass3}
    E_{j,j+1}-E_{j-1,j}=\hat\rho_j\,,
\end{equation}
We note that the anomaly-cancellation condition \eqref{eq:anomalycancel} implies $[\hat\rho_j,\hat\rho_{j'}]=0$, which means that the new Gauss laws commute with each other at different sites, ensuring that the symmetry is gaugeable. 
We also need another Gauss law imposing the quantization of $E_{j,j+1}$, namely\footnote{Note that the anomaly cancellation condition $\sum_I (n_\text{L}^{(I)})^2 =\sum_I (n_\text{R}^{(I)})^2$ implies $\sum_I(n_\text{L}^{(I)} + n_\text{R}^{(I)})=0$ mod 2, so the right hand side is a bosonic operator as it should be.}
\begin{equation}\label{eq:gaugeguass4}
\begin{aligned}
    \exp\left(2\pi i E_{j,j+1}\right)=\prod_{I=1}^{N_f}\left[\left(i(\gamma')^{(I)}_{j,j+1}\right)^{n_\text{L}^{(I)}}\left(\gamma^{(I)}_{j+1,j+2}\right)^{n_\text{R}^{(I)}}\right],
\end{aligned}
\end{equation}
where the phase factor is chosen so that $e^{4\pi i E_{j,j+1}}=1$.\footnote{The four Gauss laws above satisfy the following self-consistency condition. Taking the exponential of $2\pi i$ times \eqref{eq:gaugeguass3}, the left-hand side can be expressed in terms of the fermions using \eqref{eq:gaugeguass4}, while the right-hand side can be expressed in terms of the gauge field and fermions using \eqref{eq:gaugeguass1} and \eqref{eq:gaugeguass2}. The gauge field $A_{j,j+1}$ cancels exactly when the anomaly cancellation condition \eqref{eq:anomalycancel} is satisfied. The phase factors agree if and only if the difference between the numbers of odd left- and right-handed charges is divisible by 4, namely,
\begin{equation}
\sum_I 
{1-(-1)^{n_\text{L}^{(I)}}\over2}=\sum_I 
{1-(-1)^{n_\text{R}^{(I)}}\over2}~~\text{mod}~~4.
\end{equation}
This is indeed the case, as can be shown by an elementary number-theoretic argument. Physically, this means that the $\mathbb{Z}_2$ subgroup of the gauge group $\mathrm{U(1)}$, whose anomaly has a mod 8 classification \cite{2013NJPh...15f5002Q,2012PhRvB..85x5132R,Gu:2013azn,Kapustin:2014dxa,Tong:2019bbk,Seiberg:2023cdc,BoyleSmith:2024qgx}, is anomaly-free.}

Finally, we add the following kinetic term for the gauge fields to the Hamiltonian \eqref{eq:Hamiltonian}
\begin{equation}\label{eq:gaugekinetic}
\begin{aligned}
    \frac{e^2}{2}\sum_j\left(E_{j-1,j}+\sum_{I=1}^{N_f}\frac{n_\text{R}^{(I)}\tilde\phi_{j,j+1}^{(I)}+n_\text{L}^{(I)}\tilde\phi_{j-1,j}^{(I)}}{4\pi}\right.\\
    \left.+\sum_{I=1}^{N_f}\frac{n_\text{L}^{(I)}-n_\text{R}^{(I)}}{2\pi}\phi_j^{(I)}\right)^2.
\end{aligned}
\end{equation}
To summarize, we have constructed a general Abelian lattice gauge theory. 
The Hamiltonian is $N_f$ copies of \eqref{eq:Hamiltonian} plus \eqref{eq:gaugekinetic}, subject to the four Gauss laws Eqs.~(\ref{eq:gaugeguass1})-(\ref{eq:gaugeguass4}). 
Note that the gauge field $A_{j,j+1}$ does not show up in the Hamiltonian, since we have chosen to work with the charge density $\hat\rho$ that commutes with the Hamiltonian. 

Below, we study two special cases in detail: the Schwinger model and the fermionic 3450 gauge theory. Similar microscopic realizations of these theories were recently solved in the bosonized Villain Hamiltonian in Refs.~\cite{Seifnashri:2026ema,CSSZ}. Our lattice models differ from theirs by certain irrelevant terms, so the finite-lattice spectra are not identical, although they agree in the continuum limit. 
See also Ref. \cite{Berkowitz:2023pnz} for constructions of these models in the Euclidean modified Villain setting.

\subsection{The Schwinger model}

As a warm-up, we reproduce the spectrum of the Schwinger model, which is obtained by gauging $Q_\text{V}=Q_\text{L}+Q_{\text{R}}$ in the one-flavor fermionic Villain model. 
This microscopic realization differs from the staggered fermion model \cite{Banks:1975gq} (which was recently revived in Ref.~\cite{Dempsey:2022nys}).

Taking $N_f=1$, $n_\text{L}=n_\text{R}=1$, the Gauss laws and Hamiltonian follow from the general expressions given in Sec.~\ref{sec:chiralgauge}. 
To simplify the presentation, we conjugate the system with the unitary operator $\exp(-\frac{i}{2}\sum_j A_{j-1,j}q_{j})$. The resulting Gauss laws are the following
\begin{align}
    &i\gamma'_{j-1,j}\gamma_{j,j+1} \exp\left(i \pi  q_{j}\right)=1\,,\\
    &i\gamma_{j,j+1}\gamma'_{j,j+1}\exp\left(2\pi i w_{j,j+1}\right)=
    \exp\left(-iA_{j,j+1}\right)\,,\\
    &E_{j,j+1}-E_{j-1,j}=q_j\,,\\
    &\exp(2\pi i E_{j,j+1})=1\,.
\end{align}
The kinetic term for the gauge field is
\begin{equation}
    \frac{e^2}{2}\sum_{j}\left(E_{j-1,j}+\frac{1}{2}p_j+\frac{\tilde\phi_{j-1,j}}{2\pi}\right)^2\,.
\end{equation}
We further conjugate the system by the unitary operator $\exp\left(-2\pi i \sum_jE_{j,j+1}w_{j,j+1}\right)$. After this transformation, the gauge fields and fermions decouple, and are  completely determined by the following Gauss law constraints,
\begin{equation}\label{eq:gaugefieldguass}
    \begin{aligned}
    &\exp\left(i\pi E_{j-1,j}-i\pi  E_{j,j+1}\right)i\gamma'_{j-1,j}\gamma_{j,j+1}=1,\\
    &i\gamma_{j,j+1}\gamma'_{j,j+1}=\exp(iA_{j,j+1}),\\
    &\exp\left(2\pi i E_{j,j+1}\right)=1.
\end{aligned}
\end{equation}

Thus, the lattice Schwinger model reduces to a purely bosonic model with the following Hamiltonian,
\begin{equation}
\begin{aligned}
    H=\frac{1}{2R^2}\sum_j p_j^2&+\frac{R^2}{2}\sum_j\left(\frac{\phi_{j+1}-\phi_j}{2\pi}+w_{j,j+1}\right)^2\\
    &+\frac{e^2}{8\pi^2}\sum_j \left(\tilde\phi_{j,j+1}+\pi p_{j+1}\right)^2,
\end{aligned}
\end{equation}
and the Gauss law constraint
\begin{equation}\label{eq:schwingergauss}
    p_j+\frac{\tilde\phi_{j-1,j}-\tilde\phi_{j,j+1}}{2\pi}=0.
\end{equation} 
Indeed, this is consistent with the continuum expectation: a U(1) gauge theory coupled to a Dirac fermion is a bosonic quantum field theory which does not depend on the choice of the spin structure because fermion parity is the $\mathbb{Z}_2$ subgroup of the U(1) gauge group. 
We then perform a (bosonic) T-duality transformation \cite{Cheng:2022sgb,Seifnashri:2026ema}
\begin{equation}
\begin{aligned}
    {\cal T}_{\mathrm{boson}}:~&\phi_j\mapsto \tilde\phi_{j,j+1},
    ~~~~~&&\tilde\phi_{j,j+1}\mapsto \phi_{j+1},\\
    &q_j\mapsto w_{j,j+1},~~&& w_{j,j+1}\mapsto q_{j+1}.
\end{aligned}
\end{equation}
The Gauss law \eqref{eq:schwingergauss} becomes simply $w_{j,j+1}=0$, which means that $\phi$ is a noncompact boson. 
\begin{equation}\label{eq:schwingerHamiltonian}
\begin{aligned}
    H=\frac{R^2}{2}\sum_j p_j^2&+\frac{1}{2R^2}\sum_j\left(\frac{\phi_{j+1}-\phi_j}{2\pi}\right)^2\\
    &+\frac{e^2}{8\pi^2}\sum_j \left(\frac{\phi_j+\phi_{j+1}}{2}\right)^2.
\end{aligned}
\end{equation}
The dispersion relation, which is derived in App. \ref{app:dispersion}, is:
\begin{equation}\label{eq:schwingerdispersion}
    E(k)^2=\frac{\sin^2(k/2)}{\pi^2}+\frac{e^2R^2}{4\pi^2}\cos^2\left(k/2\right).
\end{equation}
where $k=2\pi n/N$, with $n=0,1,\cdots,N-1$. 
The continuum Hamiltonian is related to the lattice one by $H_{\text{cont}}=\frac{2\pi}{a}H$, where $a$ is a dimensionful constant that corresponds to the lattice spacing. By comparing the coefficient of the gauge field kinetic term, one finds that $e$ is related to the continuum coupling $e_{\text{cont}}$ by $e=\frac{e_{\text{cont}}a}{\sqrt{2\pi}}$. From the dispersion relation \eqref{eq:schwingerdispersion}, we find the following Schwinger mass formula
\begin{equation}
    m_{\text{Schwinger}}(R)=\frac{eR}{a}=\frac{e_{\text{cont}}R}{\sqrt{2\pi}}.
\end{equation}
In particular, at $R=\sqrt{2}$ the theory flows to the usual Schwinger model with no Thirring coupling, and we find the correct mass formula $m_{\text{Schwinger}}=e_{\text{cont}}/\sqrt{\pi}$ \cite{Lowenstein:1971fc,Casher:1974vf,Coleman:1975pw}.

\subsection{The 3450 gauge theory}\label{sec:3450gauge}

By choosing $N_f=2$ and $n_\text{L}^{(1)}=3,\ n_\text{L}^{(2)}=4,\ n_\text{R}^{(1)}=5,\ n_\text{R}^{(2)}=0$, we obtain the U$(1)_{3450}$ gauge theory. 
We keep the radii $R_1,R_2$ general; these correspond to four-fermion Thirring couplings of the Dirac fermions. 
The bosonized version of this lattice model is solved in Refs.~\cite{Seifnashri:2026ema,CSSZ}.

The Gauss laws and the Hamiltonian follow from the general expressions in Sec.~\ref{sec:chiralgauge}. We first perform a T-duality transformation to simplify the Gauss laws:
\begin{equation}
\begin{aligned}
    &\phi_j^{(1)}\mapsto -\frac{1}{2}\phi_j^{(1)}+2\phi_j^{(2)}+\frac{5}{2}\tilde\phi_{j-1,j}^{(1)}+\frac{3}{2}\tilde\phi_{j,j+1}^{(1)}+\tilde\phi_{j-1,j}^{(2)},\\
    &w_{j,j+1}^{(1)}\mapsto -\frac{1}{2}w_{j,j+1}^{(1)}+2w_{j,j+1}^{(2)}+\frac{5}{2}q_j^{(1)}+\frac{3}{2}q_{j+1}^{(1)}+q_j^{(2)},\\
    &\phi_j^{(2)}\mapsto \phi_j^{(2)}+2\tilde\phi_{j,j+1}^{(1)},~~~~~~~w_{j,j+1}^{(2)}\mapsto w_{j,j+1}^{(2)}+2q_{j+1}^{(1)},\\
    &p_j^{(1)}\mapsto -2p_j^{(1)},~~~~~~~~~~~~~~~~~p_j^{(2)}\mapsto p_j^{(2)}+4p_j^{(1)},\\
    &\tilde\phi_{j,j+1}^{(1)}\mapsto -2\tilde\phi_{j,j+1}^{(1)},~~~~~~~~~~~\tilde\phi_{j,j+1}^{(2)}\mapsto \tilde\phi_{j,j+1}^{(2)}+4\tilde\phi_{j,j+1}^{(1)},\\
    &\gamma^{(I)}_{j,j+1}\mapsto \gamma^{(I)}_{j,j+1},~~~~~~~~~~~~~~(\gamma')^{(I)}_{j,j+1}\mapsto (\gamma')^{(I)}_{j,j+1}.
\end{aligned}
\end{equation}
We then perform a unitary transformation ${\exp\left(-2\pi i\sum_j E_{j-1,j} q_j^{(1)}\right)}$ to decouple the gauge fields. Finally, we use the field redefinition ${E_{j,j+1}\mapsto E_{j,j+1}-\frac{A_{j,j+1}}{2\pi}}$ to further simplify the gauge field sector. In the end, the gauge fields and $\gamma^{(1)},(\gamma')^{(1)}$ are completely determined by the Gauss laws in \eqref{eq:gaugefieldguass} (with $\gamma,\gamma'\to \gamma^{(1)},(\gamma')^{(1)}$), 
and we are left with the following Hamiltonian,
\begin{equation}\label{eq:3450H}
\begin{split}
    H=&\frac{e^2}{8\pi^2}\sum_j \left(\phi_j^{(1)}\right)^2+\frac{2}{R_1^2}\sum_j\left(p_j^{(1)}\right)^2\\
    &+\frac{1}{2R_2^2}\sum_j\left(p_j^{(2)}+4p_j^{(1)}\right)^2\\
    &+\frac{R_1^2}{2}\sum_j\left[-\frac{\phi_{j+1}^{(1)}-\phi_{j}^{(1)}}{4\pi}+\frac{5}{2} p_{j}^{(1)}+\frac{3}{2} p_{j+1}^{(1)}\right.\\
    &\left.+ p_{j}^{(2)}+2\left(\frac{\phi_{j+1}^{(2)}-\phi_{j}^{(2)}}{2\pi}+w_{j,j+1}^{(2)}\right)\right]^2\\
    &+\frac{R_2^2}{2}\sum_j\left(\frac{\phi_{j+1}^{(2)}-\phi_{j}^{(2)}}{2\pi}+w_{j,j+1}^{(2)}+2 p_{j+1}^{(1)}\right)^2,
\end{split}
\end{equation}
and Gauss laws,
\begin{equation}\label{eq:3450gauss}
    \begin{aligned}
   &i(\gamma')^{(2)}_{j-1,j}\gamma^{(2)}_{j,j+1} \exp\left(i\pi  p^{(2)}_j+\frac{i}{2}\tilde{\phi}^{(2)}_{j-1,j}-\frac{i}{2}\tilde{\phi}^{(2)}_{j,j+1}\right)=1,\\
   &i\gamma^{(2)}_{j,j+1}(\gamma')^{(2)}_{j,j+1}e^{2\pi i w^{(2)}_{j,j+1}}=1.
\end{aligned}
\end{equation}

Therefore, the 3450 gauge theory, given by  
\eqref{eq:3450H} and \eqref{eq:3450gauss}, is equivalent to a massless Dirac fermion (with a Thirring coupling) coupled to a noncompact boson. 
The dispersion relation for the massless fermion is $E(k)= \sin(k/2)/\pi$  with $k\in \frac{2\pi}{N}\mathbb{Z}$, while that for the noncompact boson is (see App.\ \ref{app:dispersion})
\begin{equation}
\begin{aligned}
E(k)^2
&=
\frac{\sin^2(k/2)}{\pi^2}\\
&
+\frac{e^2}{\pi^2}
\left[
\frac{R_1^2}{8}(17+15\cos k)
+\frac{1}{R_1^2}
+R_2^2+\frac{4}{R_2^2}
\right].
\end{aligned}
\end{equation}
At zero spatial momentum, this gives the following continuum mass
\begin{equation}\label{eq:m3450}
    m_{3450}=e_{\text{cont}}\sqrt{\frac{2}{\pi}\left(\frac{1}{R_1^2}+4R_1^2+\frac{4}{R_2^2}+R_2^2\right)}.
\end{equation}
In the deep infrared, $\phi_j^{(1)}$ is pinned to zero, and $p_j^{(1)}$ is determined by ${0=[H,\phi_j^{(1)}]}$.\footnote{This equation is simplified by replacing $p_{j\pm1}^{(1)}$ with $p_j^{(1)}$. The resulting corrections are of higher order in the spatial momentum and are therefore irrelevant in the infrared limit.} 
A straightforward calculation shows that the infrared Hamiltonian reduces to a fermionic Villain model \eqref{eq:Hamiltonian} with radius
\begin{equation}\label{eq:RIR}
R_\text{IR} = \sqrt{4R_1^2+R_2^2\over 1+R_1^2R_2^2} .
\end{equation}
Hence, the infrared theory is a massless Dirac fermion with Thirring coupling $g_\text{IR}= {2\pi \over R_\text{IR}^2}-\pi$.

In the special case in which the ultraviolet Thirring couplings are zero, corresponding to  $R_1=R_2=\sqrt{2}$, the mass of the noncompact boson is 
\begin{equation}
m_{3450}=\frac{5e_{\text{cont}}}{\sqrt{\pi}},
\end{equation}
which matches the continuum spectrum derived in Ref.~\cite{Halliday:1985tg} and was also obtained from the bosonic Villain Hamiltonian model \cite{CSSZ}. 
The radius of the infrared Dirac fermion is $R_\text{IR}=\sqrt{2}$, which corresponds to zero Thirring coupling, $g_\text{IR}=0$. 
In other words, the deep infrared limit of the 3450 gauge theory is a free, massless Dirac fermion, consistent with the continuum result in  Ref.~\cite{Mouland:2025dct}.

\section{Outlook}

In this work, we have constructed a fermionic lattice Hamiltonian that exactly realizes the $\mathrm{U}(1)_\mathrm{L}\times \mathrm{U}(1)_\mathrm{R}$ chiral symmetry and the associated anomalies of a massless Dirac fermion in 1+1d. 
Our model is genuinely fermionic: it contains local fermionic operators and is not just a bosonized description.    
The only price paid to evade Nielsen--Ninomiya-type no-go theorems is the introduction of continuous bosonic fields, which are coupled to a Kitaev chain of Majorana fermions.  
By gauging arbitrary anomaly-free Abelian global symmetries, we obtained the most general Abelian lattice chiral gauge theories and solved the spectra in some examples.

An important future direction is to generalize the fermionic Villain construction to 3+1d. The bosonic Villain Hamiltonian of Ref.~\cite{Lu:2026jnq} realizes an exact lattice $\mathrm{U}(1)_\mathrm{V}\times\mathrm{U}(1)_\mathrm{A}$ chiral symmetry and its mixed anomaly, but its low-energy degrees of freedom are bosonic. It would be interesting to couple this model to Majorana fermions and find a local Hamiltonian invariant under the fermionic chiral symmetry operators proposed in Ref.~\cite{Thorngren:2026ydw}, in which vortex loops are decorated by Kitaev chains. Such a construction could yield a genuinely fermionic lattice model and provide a direct Hamiltonian route toward chiral fermions and chiral gauge theories in 3+1d.

Another direction is to extend the construction to Euclidean lattice models, which are more readily amenable to conventional Monte Carlo simulations. In particular, it would be interesting to formulate the fermionic Villain model within the Euclidean modified Villain framework of Refs.~\cite{Gorantla:2021svj,Sulejmanpasic:2019ytl}. 
See \cite{Berkowitz:2024iuv} for recent progress on the numerical simulation of bosonic Euclidean Villain models. 
In two spacetime dimensions, this fermionization should involve gauging an appropriate $\mathbb{Z}_2$ symmetry of the bosonic theory and coupling the resulting system to an unpaired massive Majorana fermion. The latter is the Euclidean counterpart of the Kitaev chain, and can be realized using a Majorana version of the Wilson fermion on the Euclidean lattice; see, for example, Ref.~\cite{Araki:2025xly} for a recent discussion.

\section*{Acknowledgements}

We thank 
Lei Gioia, Salvatore D. Pace, Sahand Seifnashri, and Ryan Thorngren for helpful discussions.  
We thank Shi Chen, Aleksey Cherman, Zohar Komargodski,  Sahand Seifnashri, David Tong, and Xiao-Gang Wen for comments on a draft. 
S.H.S. was supported in part by the Simons Collaboration on Ultra-Quantum Matter, which is a grant from the Simons Foundation (651444,  SHS). S.H.S. was also supported in part by the U.S. Department of Energy, Office of Science, Office of High Energy Physics of U.S. Department of Energy under grant Contract Number  DE-SC0012567. 
We thank the authors of \cite{DF} for coordinating submission of their related work on a fermionic model involving rotors and fermions.

\appendix

\section{Spectrum}\label{app:spectrum}

In this Appendix, we show that the fermionic Villain model \eqref{eq:Hamiltonian}  reproduces the continuum spectrum of a massless Dirac fermion field with a four-fermion Thirring coupling \eqref{eq:Thirring}. 
The bosonic fields are always assumed to obey the periodic boundary conditions, while the fermions can be either periodic (Ramond, $\eta=0$) or anti-periodic (Neveu-Schwarz, $\eta=1$).

\subsection{Bosonic Villain model}\label{app:bVillain}

We first recall the spectrum of the bosonic Villain Hamiltonian, which was solved exactly in Ref.~\cite{Seifnashri:2026ema}. While the Hamiltonian for the bosonic model is formally the same as \eqref{eq:Hamiltonian}, the Gauss laws are different \cite{Cheng:2022sgb,Fazza:2022fss}:
\begin{subequations}\label{eq:bgauss}
\begin{align}
 &\exp\left(2\pi i p_j+i\tilde{\phi}_{j-1,j}-i\tilde{\phi}_{j,j+1}\right)=1,\\
&\exp\left(2\pi i w_{j,j+1}\right) = 1.
\end{align}
\end{subequations}
The Hamiltonian can be diagonalized as 
\begin{equation}\label{eq:bosonHamiltonian}
\begin{aligned}
H& = {1\over 4N}\left( {Q_\text{M}\over R}+Q_\text{W}R\right)^2
+{1\over 4N}\left( {Q_\text{M}\over R}-Q_\text{W}R\right)^2\\
&
+\sum_{k}{1\over\pi}\sin({ k\over 2})\left(a_k^\dagger a_k +\frac 12\right),
\end{aligned}
\end{equation}
where $k=2\pi n/N$ with $n=1,2,\cdots, N-1$. 
The momentum and winding charges  are defined as
\begin{subequations}
\begin{align}
&Q_\text{M} = \sum_j p _j,\\
&Q_\text{W} = \sum_j w_{j,j+1}.
\end{align}
\end{subequations}
Each nonzero momentum mode is associated with creation and annihilation operators $a_k^\dagger,a_k$ satisfying $[a_k,a_{k'}^\dagger]=\delta_{k,k'}$. 
In the continuum limit, this spectrum correctly reproduces that of the $c=1$ compact boson CFT at radius $R$ (see, e.g., Ref.~\cite{Ginsparg:1988ui}).

\subsection{Fermionic Villain model}

On each link, the fermion state is determined by its eigenvalue of the fermion number operator 
\begin{equation}
n_{j,j+1}=\frac{1}{2}(1-i\gamma_{j,j+1}\gamma'_{j,j+1})=0,1.
\end{equation}
We use the first Gauss law in \eqref{eq:gauss1}
\begin{equation}
 \exp\left(i\pi  p_j+\frac{i}{2}\tilde{\phi}_{j-1,j}-\frac{i}{2}\tilde{\phi}_{j,j+1}\right)=i\gamma'_{j-1,j}\gamma_{j,j+1}
 \end{equation}
 to flip the fermion numbers on the two links $(j-1,j)$ and $(j,j+1)$. 
Hence, we can gauge fix all but one of the fermion numbers to 0. 
Without loss of generality, we choose the remaining link to be  $(0,1)$. 
The fermion number $n_{01}$ on the remaining link is determined by the eigenvalue of the fermion parity operator, i.e., $(-1)^\text{F}= 1-2n_{01}$. 
The remaining Gauss law constraint can be taken to be
\begin{equation}
\exp(i \pi Q_\text{V}) = (-1)^{\eta+1} (-1)^\text{F},
\end{equation}
where $Q_\text{V} = \sum_j p_j$. 

Next, the second Gauss law in \eqref{eq:gauss2} 
\begin{equation}
\exp(2\pi i w_{j,j+1}) = i \gamma_{j,j+1}\gamma'_{j,j+1},
\end{equation}
implies that all but one $w_{j,j+1}$ are integers. 
The last $w_{01}$ is an integer if $(-1)^\text{F}=1$ and a half-integer if $(-1)^\text{F}=-1$.
The remaining Gauss law can be taken to be
\begin{equation}
\exp(i \pi Q_\text{A}) = (-1)^\text{F},
\end{equation}
where $Q_\text{A} = 2\sum_j w_{j,j+1}$. 

Up to this point, we have completely fixed all the fermionic degrees of freedom. 
In addition, there are two remaining Gauss laws:
\begin{subequations}\label{eq:bquantization}
\begin{align}
&\exp(i\pi Q_\text{M})=(-1)^{\eta+1}(-1)^\text{F},~~~~\label{eq:bquantization1}\\
&\exp(2\pi i Q_\text{W} )= (-1)^\text{F}\,,\label{eq:bquantization2}
\end{align}
\end{subequations}
where we have defined the momentum and winding charges as $Q_\text{M} = Q_\text{V},~Q_\text{W} = \frac 12Q_\text{A}$ 
to make the comparison with the bosonic Villain model more straightforward.

The two Gauss laws in \eqref{eq:bquantization}  imply that the spectrum of the fermionic Villain model is related to that of the bosonic one via the relations in Table \ref{tab:antiperiodic}, which matches the fermionization relation in the continuum \cite{Ji:2019ugf}. 

In the continuum limit, given that the bosonic Villain model matches the spectrum of the $c=1$ compact boson CFT at radius $R$, we learn that the fermionic Villain model becomes the fermionization of the $c=1$ CFT.
The latter is known to be the massless Dirac fermion field theory with a Thirring coupling in \eqref{eq:Thirring} \cite{Coleman:1974bu,Karch:2019lnn}. 
We therefore conclude that \eqref{eq:Thirring} is the correct continuum limit of our fermionic Villain model.

\begin{table}[h]
    \centering
    \begin{tabular}{c|c|c}
         anti-periodic&$Q_\text{M}$&$Q_\text{W}$  \\
         \hline
         $(-1)^\text{F}=1$&$2\mathbb{Z}$&$\mathbb{Z}$\\
         \hline
         $(-1)^\text{F}=-1$&$2\mathbb{Z}+1$&$\mathbb{Z}+\frac{1}{2}$\\
    \end{tabular}
    ~~~~
    \begin{tabular}{c|c|c}
         periodic&$Q_\text{M}$&$Q_\text{W}$  \\
         \hline
         $(-1)^\text{F}=1$&$2\mathbb{Z}+1$&$\mathbb{Z}$\\
         \hline
         $(-1)^\text{F}=-1$&$2\mathbb{Z}$&$\mathbb{Z}+\frac{1}{2}$\\
    \end{tabular}
    \caption{Values of $Q_\text{M}$ and $Q_\text{W}$ in different sectors.}
\label{tab:antiperiodic}
\end{table}

\section{Schwinger term}\label{app:Schwinger}

Consider a left-moving Weyl fermion $\psi_\text{L}(t,x)$ in a 1+1d conformal field theory, normalized as in \eqref{eq:Thirring}. 
We focus on the left movers, but a similar discussion applies to the right movers. 
In Euclidean signature, we define the holomorphic coordinates as $z=x- i \tau$ and denote the corresponding fermion operator as $\psi(z)$. The latter obeys the OPE
\begin{equation}
\psi(z) \psi^\dagger(0) \sim {1\over 2\pi z}.
\end{equation}
The holomorphic current for U(1)$_\text{L}$ is $J(z)  = :\psi\psi^\dagger: (z)$, which obeys the OPE
\begin{equation}
J(z) J(0)\sim {1\over (2\pi)^2} {k\over z^2}.
\end{equation}
The level is $k=1$,  which captures the 't Hooft anomaly of U(1)$_\text{L}$. 

Under the Wick rotation to Lorentzian signature,  the holomorphic current leads to a Noether current whose time and spatial components are equal, i.e., $J_{\text{L},t} = J_{\text{L},x} = J(z)$.  
The equal-time commutator of the time-component of the U(1)$_\text{L}$ Noether current follows from the Euclidean OPE above:
\begin{equation}
\begin{aligned}
    &[J_{\text{L},t}(0,x),J_{\text{L},t}(0,0)]={1\over 4\pi^2}\lim_{\epsilon\rightarrow 0}\left[\frac{1}{(x+i\epsilon)^2}-\frac{1}{(x-i\epsilon)^2}\right]\\
    &
    ={ i\over2\pi} \delta'(x).
    \end{aligned}
\end{equation}
This matches the continuum limit of our lattice result in \eqref{eq:LRSchwinger}.

\section{Spectral flow}\label{app:spectralflow}

The spectral flow formula relates the conformal weights $(h,\bar h)$ and the symmetry charges $Q_\text{L},Q_\text{R}$ of states in the sector twisted by a  U(1)$_\text{L}$ rotation of angle $\theta$  to those in the untwisted sector (i.e., $\theta=0$) \cite{Schwimmer:1986mf}:
\begin{equation}
\begin{aligned}
&h^\theta = h - {\theta\over 2\pi}Q_\text{L} +\frac k2 \left({\theta\over 2\pi}\right)^2,~~
&& Q_\text{L}^\theta
=Q_\text{L}-{\theta\over 2\pi} k ,\\
&\bar h^\theta = \bar h,~~&& Q_\text{R}^\theta = Q_\text{R}
\end{aligned}
\end{equation}
with $k=1$ for the Dirac fermion field theory in \eqref{eq:Thirring}. 
By a conformal map, the twisted sector states are mapped to point operators living at the end of the U(1)$_\text{L}$ symmetry defect line. 
Starting with the identity operator with $h=\bar h=0$ and $Q_\text{L}=Q_\text{R}=0$, a $2\pi$ spectral flow by U(1)$_\text{L}$ maps the identity to the left-moving Weyl fermion field $\psi_\text{L}$  operator with $h^{2\pi} = 1/2, Q_\text{L}^{2\pi}=-1,Q_\text{R}^{2\pi} = 0$.

\section{More on the six-fermion deformation}\label{app:SMG}

The ungauged 3450 model is defined as two copies of the fermionic Villain Hamiltonian \eqref{eq:Hamiltonian} deformed by the six-fermion interactions in \eqref{eq:sixfermion}. In this Appendix, we give a more explicit demonstration that these deformations generate a symmetric mass gap and drive the ungauged 3450 model into a trivially gapped phase.

To simplify the analysis, we first perform the fermionic T-duality transformation in \eqref{eq:SMGTduality}, followed by another one as in \eqref{eq:fermionTduality}. In the new duality frame, where all fields are capitalized, the Hamiltonian becomes
\begin{equation}\label{eq:deformHamiltonian}
\begin{split}
    H=&\frac{1}{2R_1^2}\sum_j\left(P_j^{(1)}\right)^2+\frac{1}{2R_2^2}\sum_j\left(P_j^{(2)}\right)^2\\
    &+\frac{R_1^2}{2}\sum_j\left(W_{j,j+1}^{(1)}+P_{j+1}^{(2)}+\frac{\Phi_{j+1}^{(1)}-\Phi_{j}^{(1)}}{2\pi}\right)^2\\
    &+\frac{R_2^2}{2}\sum_j\left(W_{j,j+1}^{(2)}-P_{j}^{(1)}+\frac{\Phi_{j+1}^{(2)}-\Phi_{j}^{(2)}}{2\pi}\right)^2\\
    &-\lambda\sum_j\cos\left(2\Phi_j^{(1)}\right)-\lambda\sum_j\cos\left(2\Phi_j^{(2)}\right).
\end{split}
\end{equation}
Note that in the new duality frame, the deformation terms (with coefficient $\lambda$)  become simple and decoupled, while the original Hamiltonian becomes more complicated. 
This Hamiltonian can be equivalently written as
\begin{equation}
\begin{aligned}
    H&=\frac{1}{2\kappa_1}\sum_j\left(P_j^{(1)}-\kappa_1R_2^2\left(W_{j,j+1}^{(2)}+\frac{\Phi_{j+1}^{(2)}-\Phi_{j}^{(2)}}{2\pi}\right)\right)^2\\
    &+\frac{1}{2\kappa_2}\sum_j\left(P_j^{(2)}+\kappa_1R_2^2\left(W_{j-1,j}^{(1)}+\frac{\Phi_{j}^{(1)}-\Phi_{j-1}^{(1)}}{2\pi}\right)\right)^2\\
    &+\frac{\kappa_1}{2}\sum_j\left(W_{j,j+1}^{(1)}+\frac{\Phi_{j+1}^{(1)}-\Phi_{j}^{(1)}}{2\pi}\right)^2\\
    &+\frac{\kappa_2}{2}\sum_j\left(W_{j,j+1}^{(2)}+\frac{\Phi_{j+1}^{(2)}-\Phi_{j}^{(2)}}{2\pi}\right)^2\\
    &-\lambda\sum_j\cos\left(2\Phi_j^{(1)}\right)-\lambda\sum_j\cos\left(2\Phi_j^{(2)}\right),
\end{aligned}
\end{equation}
where we have defined ${\kappa_1=\frac{R_1^2}{1+R_1^2R_2^2}}$, ${\kappa_2=\frac{R_2^2}{1+R_1^2R_2^2}}$. 
The Gauss laws are preserved by these duality transformations:
\begin{align}
&\exp(i \pi Q_j^{(I)})=i (\Gamma')_{j-1,j}^{(I)}\Gamma^{(I)}_{j,j+1},~~~~~\label{eq:deformGauss1}\\
&\exp(2\pi i W^{(I)}_{j,j+1}) = i \Gamma^{(I)}_{j,j+1}(\Gamma')^{(I)}_{j,j+1}\label{eq:deformGauss2}\,,
\end{align}
where $Q_j^{(I)} = P_j^{(I)} +  {\tilde\Phi_{j-1,j} - \tilde\Phi_{j,j+1}\over 2\pi}$.

In the limit $\lambda\rightarrow\infty$, the $P^{(1)}$ term combines with the $\cos2\Phi^{(1)}$ term to form a harmonic oscillator, whose energy gap  is proportional to $\sqrt{\lambda}$. The same is true for the $P^{(2)}$ term and $\cos2\Phi^{(2)}$ term. We have the following low-energy effective Hamiltonian
\begin{equation}
\begin{split}
    H_{\rm{eff}}&=\frac{\kappa_1}{2}\sum_j\left(W_{j,j+1}^{(1)}+\frac{\Phi_{j+1}^{(1)}-\Phi_{j}^{(1)}}{2\pi}\right)^2\\
    &+\frac{\kappa_2}{2}\sum_j\left(W_{j,j+1}^{(2)}+\frac{\Phi_{j+1}^{(2)}-\Phi_{j}^{(2)}}{2\pi}\right)^2
\end{split}
\end{equation}
with each $\Phi_j^{(I)}$ pinned to a multiple of $\pi$ by the cosine potentials.

We then use \eqref{eq:deformGauss1} to set the fermion number $\frac{1}{2}\left(1-i\Gamma^{(I)}_{j,j+1}(\Gamma')^{(I)}_{j,j+1}\right)$ to be 0 on every link except link $(0,1)$. The residual Gauss laws are
\begin{align}
    &\exp(2 \pi i Q_j^{(I)})=1,\label{eq:deformGauss3}\\
    &\exp\left(i\pi\sum_j P_j^{(I)}\right)=(-1)^{\eta+1}(-1)^{\text{F}^{(I)}},\label{eq:deformGauss4}
\end{align}
together with \eqref{eq:deformGauss2}. Here $\eta=0(1)$ for (anti-)periodic boundary condition. 
Next, we choose a gauge so that the operators $W_{j,j+1}^{(I)}$ are diagonalized. For example, one can use \eqref{eq:deformGauss3} to set $W_{j,j+1}^{(I)}=0$ for every $j\ne 0$. 

The ground state is then given by $W_{0,1}^{(I)}=0$, and $\Phi^{(I)}_j=m^{(I)}\pi$ for all $j$ and some integers $m^{(I)}$. Different choices of $m^{(I)}$ are related by the residual Gauss law \eqref{eq:deformGauss4}, and we conclude that the ground state is gapped and non-degenerate.

\section{Dispersion relations}\label{app:dispersion}

We first derive the dispersion relation of the Schwinger model \eqref{eq:schwingerHamiltonian}. We introduce fields in the momentum space as
\begin{equation}\label{eq:momentumfield}
\phi_j=\frac{1}{\sqrt N}\sum_k e^{ikj}\phi_k,
\qquad
p_j=\frac{1}{\sqrt N}\sum_k e^{ikj}p_k,
\end{equation}
where $k=\frac{2\pi n}{N}$ and $n=0,1,\cdots,N-1$. 
The Hamiltonian becomes
\begin{equation}
\begin{aligned}
    H&=\frac{R^2}{2}\sum_kp_{-k}p_k\\
    &+\sum_k \left(\frac{1}{2\pi^2R^2}\sin^2(k/2)+\frac{e^2}{8\pi^2}\cos^2(k/2)\right)\phi_{-k}\phi_k.
\end{aligned}
\end{equation}
Using the Heisenberg equations $\dot\phi_k=i[H,\phi_k]$, $\dot p_k=i[H,p_k]$ and taking all fields to have time dependence $e^{-iEt}$, we derive the dispersion relation in \eqref{eq:schwingerdispersion}.

Next, we derive the dispersion relation of the 3450 gauge theory.
To determine the oscillator spectrum, we can ignore the integer-valued field $w_{j,j+1}^{(2)}$, which only affects the zero modes. 
For each copy of fields, we use the momentum space version as introduced in \eqref{eq:momentumfield}, and the Hamiltonian becomes
\begin{align}\label{eq:dispersionHamiltonian}
&H=\sum_k\bigg\{
\frac{e^2}{8\pi^2}\phi_{-k}^{(1)}\phi_k^{(1)}
+\frac{2}{R_1^2}p_{-k}^{(1)}p_k^{(1)}
+\frac{1}{2R_2^2}
\left|p_k^{(2)}+4p_k^{(1)}\right|^2
\nonumber\\
&+\frac{R_1^2}{2}
\left|
-\frac{e^{ik}-1}{4\pi}\phi_k^{(1)}
+\frac{5+3e^{ik}}{2}p_k^{(1)}
+p_k^{(2)}
+\frac{e^{ik}-1}{\pi}\phi_k^{(2)}
\right|^2
\nonumber\\
&+\frac{R_2^2}{2}
\left|
\frac{e^{ik}-1}{2\pi}\phi_k^{(2)}
+2e^{ik}p_k^{(1)}
\right|^2
\bigg\}.
\end{align}
The determinant from the resulting Heisenberg equations factorizes as
\begin{equation}
\begin{aligned}
0={}&
\left[
E^2-\frac{\sin^2 (k/2)}{\pi^2}
\right]
\bigg[
E^2-\frac{\sin^2 (k/2)}{\pi^2}\\
&
-\frac{e^2}{\pi^2}
\left(
\frac{R_1^2}{8}(17+15\cos k)
+\frac{1}{R_1^2}
+R_2^2+\frac{4}{R_2^2}
\right)
\bigg].
\end{aligned}
\end{equation}
This gives the two dispersion relations reported in Sec.\ \ref{sec:3450gauge}. 
Note that the massive branch never has a zero.

In taking the continuum limit, we set
\begin{equation}
e=\frac{e_{\rm cont}a}{\sqrt{2\pi}},\qquad
E=\frac{a}{2\pi}E_{\rm cont},\qquad
k=k_{\rm cont}a,
\end{equation}
and restrict to $|k|\ll 1$. In this limit, $\cos k\to 1$, and the dispersion relation reduces to:
\begin{equation}
    E_{\rm cont}^2=k_{\rm cont}^2+\frac{2}{\pi}e_{\rm{cont}}^2\left(4R_1^2+\frac{1}{R_1^2}+R_2^2+\frac{4}{R_2^2}\right),
\end{equation}
which is the standard dispersion relation for a massive boson in the continuum with mass given in \eqref{eq:m3450}.

In the deep infrared, the massive boson decouples and we are left with a massless Dirac fermion with its Thirring coupling $g_\text{IR}$ determined by \eqref{eq:RIR}, as discussed in Sec.\ \ref{sec:3450gauge}. 
This can also be derived from the Hamiltonian in momentum space in \eqref{eq:dispersionHamiltonian}. 
Setting $k=0$, the Hamiltonian becomes
\begin{equation}
    \begin{aligned}
        H_{k=0}
        &=\frac{e^2}{8\pi^2}\left(\phi_0^{(1)}\right)^2+\frac{1+R_1^2R_2^2}{2(R_2^2+4R_1^2)}\left(p_0^{(2)}\right)^2\\
        &+\frac{2(R_2^2+4R_1^2)(1+R_1^2R_2^2)}{R_1^2R_2^2}\left(p_0^{(1)}+\frac{R_1^2}{R_2^2+4R_1^2}p_0^{(2)}\right)^2. 
    \end{aligned}
\end{equation}
The infrared limit corresponds to taking $e$ to be large, which pins ${\phi^{(1)}_0=0}$. Then ${p_0^{(1)}}$ is fixed to be ${-\frac{R_1^2}{R_2^2+4R_1^2}p_0^{(2)}}$ by demanding ${0=[H_{k=0},\phi^{(1)}_0]}$.
We are therefore left with the zero mode of a fermionic Villain model in \eqref{eq:Hamiltonian}
\begin{equation}
{1\over 2R_\text{IR}^2 } (p_0^{(2)})^2
\end{equation}
with $R_\text{IR}$ given in \eqref{eq:RIR}.

\bibliographystyle{ytphys}
\bibliography{ref}

\end{document}